\documentclass[12pt,epsf]{article}
\usepackage{graphicx}
\usepackage{amssymb,amsmath}

\begin{document}
%%%%%%%%%%%%%%%%%%%%%%%%%%%%%%%%%%%%%%%%%%%

\def\a{\alpha}
\def\b{\beta}
\def\c{\varepsilon}
\def\d{\delta}
\def\e{\epsilon}
\def\f{\phi}
\def\g{\gamma}
\def\h{\theta}
\def\k{\kappa}
\def\l{\lambda}
\def\m{\mu}
\def\n{\nu}
\def\p{\psi}
\def\q{\partial}
\def\r{\rho}
\def\s{\sigma}
\def\t{\tau}
\def\u{\upsilon}
\def\v{\varphi}
\def\w{\omega}
\def\x{\xi}
\def\y{\eta}
\def\z{\zeta}
\def\D{\Delta}
\def\G{\Gamma}
\def\H{\Theta}
\def\L{\Lambda}
\def\F{\Phi}
\def\P{\Psi}
\def\S{\Sigma}
\def\BR{{\rm Br}}
\def\o{\over}
\def\beq{\begin{eqnarray}}
\def\eeq{\end{eqnarray}}
\newcommand{\nn}{\nonumber \\}
\newcommand{\gsim}{ \mathop{}_{\textstyle \sim}^{\textstyle >} }
\newcommand{\lsim}{ \mathop{}_{\textstyle \sim}^{\textstyle <} }
\newcommand{\vev}[1]{ \left\langle {#1} \right\rangle }
\newcommand{\bra}[1]{ \langle {#1} | }
\newcommand{\ket}[1]{ | {#1} \rangle }
\newcommand{\EV}{ {\rm eV} }
\newcommand{\KEV}{ {\rm keV} }
\newcommand{\MEV}{ {\rm MeV} }
\newcommand{\GEV}{ {\rm GeV} }
\newcommand{\TEV}{ {\rm TeV} }
\def\diag{\mathop{\rm diag}\nolimits}
\def\Spin{\mathop{\rm Spin}}
\def\SO{\mathop{\rm SO}}
\def\O{\mathop{\rm O}}
\def\SU{\mathop{\rm SU}}
\def\U{\mathop{\rm U}}
\def\Sp{\mathop{\rm Sp}}
\def\SL{\mathop{\rm SL}}
\def\tr{\mathop{\rm tr}}

%added by FT
\newcommand{\bear}{\begin{array}}  
\newcommand {\eear}{\end{array}}
\newcommand{\la}{\left\langle}  
\newcommand{\ra}{\right\rangle}
\newcommand{\non}{\nonumber}  
\newcommand{\ds}{\displaystyle}
\newcommand{\red}{\textcolor{red}}
\def\ubl{U(1)$_{\rm B-L}$}
\def\REF#1{(\ref{#1})}
\def\lrf#1#2{ \left(\frac{#1}{#2}\right)}
\def\lrfp#1#2#3{ \left(\frac{#1}{#2} \right)^{#3}}
\def\OG#1{ {\cal O}(#1){\rm\,GeV}}

%%%%%%%%%%%%%%%%%%%%%%%%%%%%%%%
%%%    remove the following commands when finalizing
%%%%%%%%%%%%%%%%%%%%%%%%%%%%%%%
\def\TODO#1{ {\bf ($\clubsuit$ #1 $\clubsuit$)} }
%%%%%%%%%%%%%%%%%%%%%%%%%%%%%%%
%%%%%%%%%%%%%%%%%%%%%%%%%%%%%%%

%%%%%%%%%%%%%%%%%%%%%%%%%%%%%%%%%%%%%%%%%%%%%%%%%%%%%%%%%%%%%%%%%%%%

\baselineskip 0.7cm

\begin{titlepage}

\begin{flushright}
UT-12-09 \\
IPMU12-0075
\end{flushright}

\vskip 1.35cm
\begin{center} 
{\large \bf 
Peccei-Quinn extended gauge-mediation model\\ with vector-like matter and 125GeV Higgs
}
\vskip 1.2cm

{Kazunori Nakayama$^{a,b}$ and  Norimi Yokozaki$^{a,b}$}

\vskip 0.4cm

{\it
$^a$Department of Physics, University of Tokyo, Tokyo 113-0033, Japan\\
$^b$Kavli Institute for the Physics and Mathematics of the Universe,
University of Tokyo, Kashiwa 277-8583, Japan\\
}

\vskip 1.5cm

\abstract{

We construct a gauge-mediated SUSY breaking model with vector-like matters combined with the Peccei-Quinn mechanism 
to solve the strong CP problem.
The Peccei-Quinn symmetry plays an essential role for generating sizable masses for the vector-like matters
and the $\mu$-term without introducing dangerous CP angle.
The model naturally explains both the 125GeV Higgs mass and the muon anomalous magnetic moment.
The stabilization of the Peccei-Quinn scalar and the cosmology of the saxion and axino are also discussed.

 }
\end{center}
\end{titlepage}

\setcounter{page}{2}

%%%%%%%%%%%%%%%%%%%%%%%%%%%%%%%%%
\section{Introduction}
%%%%%%%%%%%%%%%%%%%%%%%%%%%%%%%%%

Recently ATLAS and CMS collaborations have discovered the Higgs boson at the mass around 125 GeV~\cite{ATLAS_Higgs1, CMS_Higgs1}. 
%Moreover, ATLAS excluded the Higgs boson lighter than 117.5\,GeV~\cite{ATLAS:Higgs}. 
In conventional gauge mediated supersymmetry (SUSY) breaking (GMSB) models~\cite{Giudice:1998bp}, 
it is difficult to explain the Higgs mass of $124$--$126$\,GeV, unless the SUSY particles are as heavy as $10$--$100$\,TeV. 
Obviously such heavy SUSY particles are not favored from a viewpoint of naturalness. 
Moreover, there is another possible indication of TeV scale SUSY particles : the muon anomalous magnetic moment (muon $g-2$). 
In fact the experimental value of the muon $g-2$ is deviated from the SM prediction at about 3 $\sigma$ level~\cite{Hagiwara:2006jt,Davier:2009ag}.
This deviation can be naturally explained with TeV scale SUSY particles and relatively large $\tan\beta$.

One of the easiest way to raise the Higgs mass up to 125\,GeV while keeping the SUSY particle masses around 1\,TeV
is to introduce additional vector-like matter~\cite{Moroi:1991mg}, 
which couple to the minimal SUSY standard model (MSSM) Higgs via yukawa interactions.
Similarly to the top yukawa coupling which radiatively increases the Higgs mass, the new yukawa coupling
gives an additional potential for the Higgs fields and the Higgs can be heavier than the MSSM case.
In order to avoid the gauge anomaly, the vector-like matter may be in the ${\bf 10}$ and $\overline{\bf 10}$ representations of
SU(5) grand unified theory (GUT) gauge group.
This type of extension of the MSSM has been recently discussed in Refs.~\cite{Asano:2011zt, Endo:2011mc, Moroi:2011aa, Endo:2011xq}
in the light of recent ATLAS and CMS results.
In particular, it was pointed out that the 125\,GeV Higgs and the muon $g-2$ can be explained simultaneously in this class of models~\cite{Endo:2011mc, Endo:2011xq}.

From a viewpoint of model building, however, this is far from complete.
First, the $\mu/B\mu$-problem in GMSB must be solved in order to obtain a correct electroweak symmetry breaking (EWSB) minimum.
We cannot discuss the SUSY CP problem unless the mechanism for generating $\mu/B\mu$ is specified.\footnote{In the gravity-mediation models, $\mu$ and $B\mu$ terms as well as the SUSY masses of the vector-like matter can be generated~\cite{Asano:2011zt} 
by the Giudice-Masiero mechanism~\cite{Giudice:1988yz}.}
%%%%%%%%%%%%%%%
Second, the SUSY masses of vector-like matter, which can in principle take arbitrary values,
must also happen to be around the weak scale in order to raise the Higgs mass.
Moreover, in general there can be both the up-type and down-type Higgs couplings to the vector-like matter,
the latter of which tends to decrease the Higgs mass.
One must somehow tune the latter coupling so as not to affect the Higgs potential.
Finally, the strong CP problem was not addressed in these frameworks.

In this paper we consider the extended GMSB model with Peccei-Quinn (PQ) symmetry~\cite{Peccei:1977hh}
in order to deal with above mentioned problems.
MSSM fields as well as vector-like matter are charged under the U(1)$_{\rm PQ}$, so that the $\mu$-term and
the mass terms for the vector-like matter are forbidden.
Then they are generated by the vacuum expectation value (VEV) of the PQ scalar field
at a correct scale for the PQ scale of $10^{9}$--$10^{11}$\,GeV~\cite{Kim:1983dt}.
The spontaneous PQ symmetry breaking predicts an almost massless Nambu-Goldstone boson, called axion,
which dynamically solves the strong CP problem~\cite{Peccei:1977hh,Kim:1986ax}.
Dangerous CP angle is not introduced, hence it also solves the $\mu/B\mu$-problem and the SUSY CP problem.
Unwanted couplings between down-type Higgs and the vector-like matter are forbidden by the PQ symmetry.

This paper is organized as follows.
In Sec.~\ref{sec:model} our model is introduced, and it is shown that the Higgs mass of 125\,GeV and
the muon $g-2$ can be explained simultaneously.
In Sec.~\ref{sec:stab} a mechanism for stabilizing the PQ scalar is described.
Cosmological constraints on our model are also discussed.
Sec.~\ref{sec:conc} is devoted for summary and conclusions.

%Recently, some attempts to extend GMSB models are done, which is consistently with muon $g-2$ observation. 
%The one of the motivated way of the extension is to introduce the vector-like matter~\cite{Asano:2011zt,Endo:2011mc}.

%SUSY CP problem is possibly solved depending on the origin of $\mu$ and $B\mu$ term.

%In this paper, we consider the origin of $\mu$ term and the SUSY mass for the vector-like matter. The Higgs B-term is generated by renormalization group evolution, therefore its size is correct and no CP violating phase arises from the Higgs sector.

%%%%%%%%%%%%%%%%%%%%%%%%%%%%%%%%%%%%%%%%
\section{The extended GMSB model with PQ symmetry}  \label{sec:model}
%%%%%%%%%%%%%%%%%%%%%%%%%%%%%%%%%%%%%%%%

We consider a model of gauge-mediated SUSY breaking with global U(1)$_{\rm PQ}$ symmetry.
The superpotential of the model is given by
\begin{eqnarray}
W &=& W_{\rm MSSM} + W_{\rm ext} +  W_{\rm PQ+mess}, 
\end{eqnarray}
where $W_{\rm MSSM}$ contains Yukawa interaction terms in the MSSM (except for the $\mu$-term),
$W_{\rm PQ+mess}$ is the superpotential for the PQ  and messenger sector, 
and $W_{\rm ext}$ consists of the additional vector-like matter.
They are given by
\begin{equation}
	W_{\rm ext} = \lambda_1 \frac{\phi^2}{M_P} H_u H_d + \frac{\phi^2}{M_P} (\lambda_2 Q' \bar{Q}'+ \lambda _3 \bar{U}' {U}' + 	\lambda_4 \bar{E}' E' ) + Y' Q' H_u \bar{U}', \label{eq:wext}
\end{equation}
and
\begin{equation}
	W_{\rm PQ+mess} = k \phi \Psi_{\rm PQ} \overline \Psi_{\rm PQ} + \kappa X \Psi_{\rm mess}\overline{\Psi}_{\rm mess}, \label{eq:pq} 
\end{equation}
where $\phi$ is a PQ symmetry breaking field and $(Q' , \bar{U}' , \bar{E}')$ and $(\bar{Q}' , {U}' , {E}')$ are the extra vector-like matter, which transform ${\bf 10}$ and $\overline{{\bf 10}}$ under the $SU(5)$ GUT gauge group, respectively,
and $M_P$ is the reduced Planck scale.
The PQ quarks, $\Psi_{\rm PQ}$ and $\overline \Psi_{\rm PQ}$, 
and the messenger fields, $\Psi_{\rm mess}$ and $\overline \Psi_{\rm mess}$
transform as ${\bf 5}$ and $\overline {\bf 5}$ under the SU(5) GUT.
The spurion field $X$ gives the SUSY breaking mass to the messenger.
The PQ scalar $\phi$ obtains a VEV of order of $10^{10}-10^{12}$\,GeV as explained in the next section.
Thus it spontaneously breaks the U(1)$_{\rm PQ}$ symmetry and the associated NG boson behaves as the axion
which solves the strong CP problem~\cite{Kim:1979if,Dine:1981rt}.
Then the SUSY masses for the Higgs and vector-like matter are generated as $\mu \sim M_{Q', U', E'}  \sim \lambda_i  \langle\phi\rangle^2/M_P$ for $\left< \phi\right> \simeq 10^{10}-10^{12}$ GeV.
The PQ charge assignments on these fields are summarized in Table.~\ref{table:charge}.

%%%%%%%%%%%%%%%%%%%%%%%%%%%%%%%%%%%%
%%%%%%%%%%%%%%%%%%%%%%%%%%%%%%%%%%%%%%%%%%%%%
\begin{table}[t!]
  \begin{flushleft}
    \begin{tabular}{  c | c | c | c | c | c | c | c | c  }
%     \hline 
         ~          &  $H_u$ & $H_d$ & $\phi$ & $\overline{{\bf 5}}_{\rm M}$ & ${\bf 10_M}$ & ${\bf 10}'$ & $\overline{\bf 10}'$  \\
       \hline \hline
         $U(1)_{\rm PQ}$  & $q_H$  & $-2-q_H$ & $1$  & $2+3q_H/2$   & $-q_H/2$   & $-q_H/2$ & $-2+q_H/2$       \\ \hline 
       R$_P$ &$+$ & $+$ & $+$  & $-$ & $-$ & $+$ & $+$ &  \\ \hline
    \end{tabular}
    %%%%%%%%%%%%%%%%%%%%%%%%%%%%%
    
    \vspace{10pt}
        \begin{tabular}{  c | c | c | c | c | c | c  }
%     \hline 
         ~          & $\Psi_{\rm PQ}$ & $\overline{\Psi}_{\rm PQ}$ & $X$ & $\Psi_{\rm mess}$ & $\overline\Psi_{\rm mess}$ \\
       \hline \hline
         $U(1)_{\rm PQ}$  & $q_{\Psi}$  & $-1 - q_{\Psi}$ & $0$  & $q_{\rm mess}$ & $-q_{\rm mess}$      \\ \hline 
       R$_P$ &$+$ & $+$ & $+$  & $+$ & $+$ &   \\ \hline
    \end{tabular}
    \caption{ 		
     	Charge assignments on chiral superfields fields in the model under 
	the $U(1)_{\rm PQ}$ and R-parity ($+$ : even, $-$ : odd). 
	$\overline{\bf 5}_{\rm M}$ and ${\bf 10_M}$ are the MSSM matter fields.
     }
  \label{table:charge}
  \end{flushleft}
\end{table}
%%%%%%%%%%%%%%%%%%%%%%%%%%%%%%%%%%%%%%%%%%%%%

Couplings between the Higgs and the extra matters give large radiative corrections to the lightest Higgs mass. 
The last term in Eq.~(\ref{eq:wext}), $Y' Q' H_u \bar{U}'$, gives positive corrections to the Higgs mass squared as~\cite{Babu:2008ge, Martin:2009bg}~\footnote{The finite corrections, which arise through trilinear couplings, $A' Y' \bar{Q'} H_u \bar{U}'$ are not shown here. In numerical calculations, they are included.}
\begin{eqnarray}
\Delta m_{h^0}^2 \simeq \frac{3 v^2}{4\pi^2} {Y'}^4 \sin^4\beta \left[ \ln\frac{M_S^2}{M_F^2} -\frac{1}{6} \left(1-\frac{M_F^2}{M_S^2}\right)\left(5 - \frac{M_F^2}{M_S^2} \right) \right]
\end{eqnarray}
where $M_F (\sim M_{Q', U'})$ is a fermionic mass of the extra matter and $M_S$ is the average of the scalar masses, $M_S^2=M_{Q', U'}^2 + m_{\rm soft}^2$ where $m_{\rm soft}$ denotes the SUSY braking contribution. 
Here we consider the decoupling limit. 
On the other hand, if the coupling like ${Y}'' \bar{Q}' H_d U'$ would exist, it gives negative contributions to the Higgs mass squared, which can be significantly large for ${Y''} \sim1$ and large $\mu$ parameter. 
The correction is given by~\cite{Martin:2009bg, Endo:2011mc}
\begin{eqnarray}
\Delta m_{h^0}^2 \simeq -\frac{3 v^2}{4\pi^2} {Y''}^4 \sin^4\beta \frac{\mu^4}{12 M_S^4},
\end{eqnarray}
where corrections suppressed by $\tan\beta$ are neglected. Moreover, a CP violating phase exists, since phases of $M_{Q'}$, $M_{U'}$, $Y'$ and $Y''$ can not be removed simultaneously.
In our model, $Y''$ is of the order of $\left<\phi\right>^6/M_P^6$ thanks to the PQ symmetry and hence highly suppressed. 
Therefore this effect is negligibly small.  Note that there are no significant corrections to couplings, $h$-$gg$ and $h$-$\gamma \gamma$, as long as $Y''$ is negligibly small; the corrections are approximately proportional to $\Delta=\sum_{i=1,2} (\partial \log M_{F,i}^2)/(\partial \log v_u)$ (see e.g.,~\cite{Low:2009di}), where $M_{F,i}$ is the eigenvalue of the fermion mass matrix of the extra-matters, 
and $\Delta$ vanishes for $Y''=0$.
%where $M_S$ is an average scalar mass for the extra matters, given by $M_S^2=M_{Q', U', E'}^2 + m_{\rm soft}^2$. 
%Here we assume the decoupling limit. 
%Since $Y''$ is highly suppressed as $\left<\phi^6\right>/M_P^6$ in our model, the effect is negligibly small.

Due to the additional contributions, the Higgs mass ($m_h$) can be easily around 125\,GeV with neither the enhanced trilinear coupling of the stop nor heavy stops.
This is welcome in terms of the muon $g-2$. In fact, the experimental value of the muon $g-2$ is deviated from SM prediction at 3.2$\sigma$~\cite{Hagiwara:2006jt}:
\begin{eqnarray}
a_\mu^{\rm exp}- a_\mu^{\rm SM} = (26.1 \pm 8.0 ) \times 10^{-10}.
\end{eqnarray}
The deviation can be naturally explained for large $\tan\beta$ and relatively small soft SUSY breaking mass. 
In the MSSM without large $A_t$, $m_h \simeq 125$ GeV requires the stop mass of $\mathcal{O}(10)$ TeV. Consequently, the Higgs mass and the muon $g-2$ can not be explained simultaneously.
Remarkably, in the model with the vector-like matter, the deviation can be explained consistently with the Higgs mass $m_h \simeq 125$\,GeV in GMSB due to the additional contributions to the Higgs mass~\cite{Endo:2011mc, Endo:2011xq}. 
In the numerical calculation, the SUSY mass spectrum is calculated by Suspect package~\cite{Djouadi:2002ze}, which is modified to include 2-loop renormalization group equations for the extra matter. 
The Higgs mass and the muon $g-2$ are evaluated by FeynHiggs package~\cite{Hahn:2010te}.

\begin{figure}
\begin{center}
\includegraphics[width=15cm]{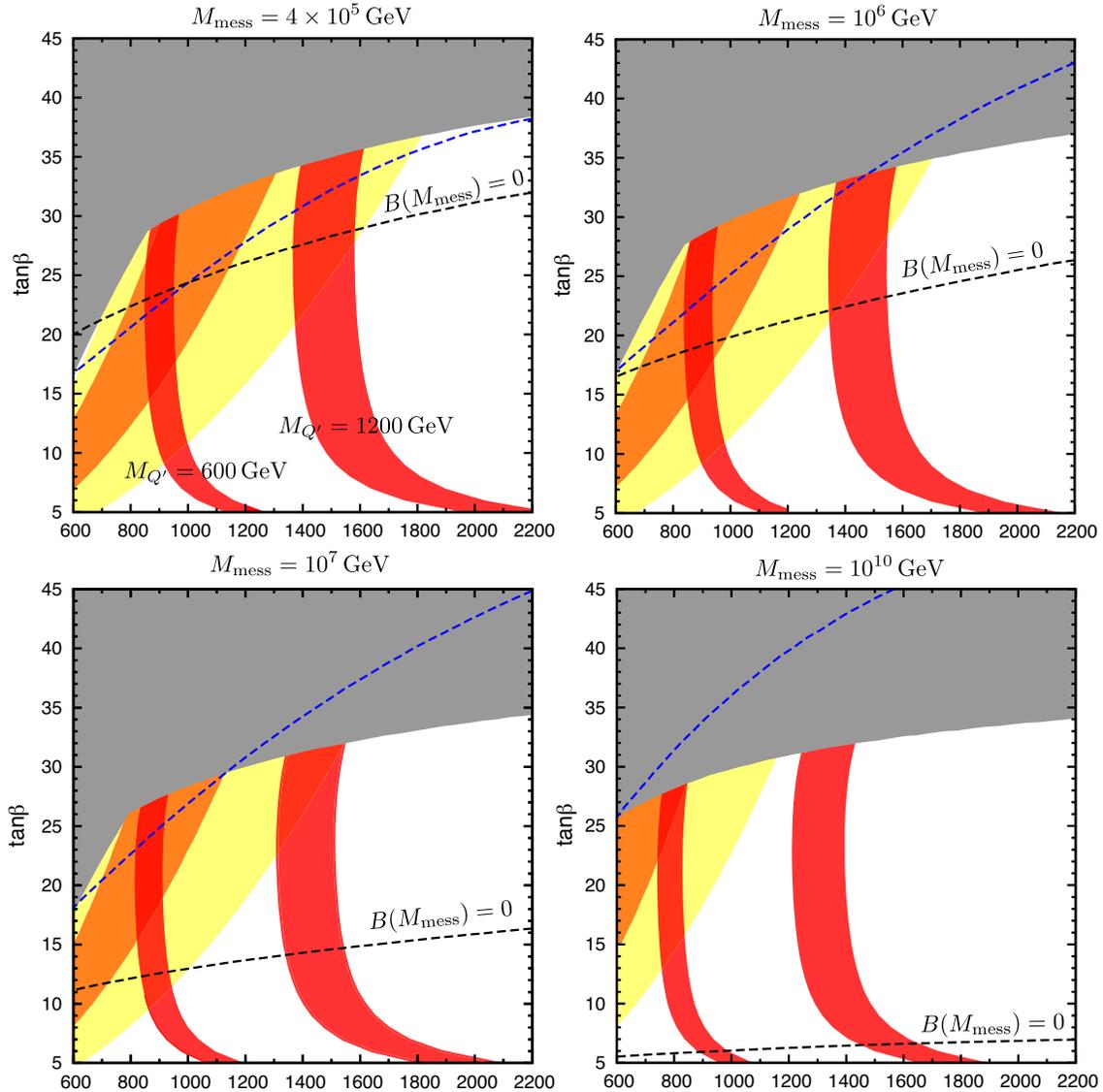}
\caption{The contours of the Higgs mass and muon $g-2$ on $m_{\tilde{g}}$-$\tan\beta$ plane for different messenger scales. The horizontal axis displays the gluino mass in the unit of GeV.
On the blue dashed line, the lightest neutralino mass and the lightest stau mass are equal, $m_{\tilde{\tau}_1}=m_{\chi_0}$. 
In the red bands, the Higgs mass of $124\, {\rm GeV} < m_h < 126\, {\rm GeV}$ 
is explained for $M_{Q', U'}=600$ GeV and $1200$ GeV.
The gray region is excluded by the vacuum stability bound and/or the bound from OPAL experiment~\cite{Abbiendi:2005gc}.}
\label{fig:higgs_gm2}
\end{center}
\end{figure}

In Fig.~{\ref{fig:higgs_gm2}}, contours of the Higgs mass and muon $g-2$ for different messenger scales are shown. 
The gray regions are excluded by the constraint from the charge breaking vacuum~\cite{Hisano:2010re} (see Ref.~\cite{Endo:2012rd} for GMSB with vector-like matter) or/and the LEP bound on the stau mass, $m_{\tilde{\tau}}< 87.4$ GeV~\cite{Abbiendi:2005gc}.  
Due to additional negative corrections to $m_{H_u}^2$ from the vector-like matter, the predicted value of $\mu$ parameter tends to be larger than that of MSSM. As a result, a trilinear coupling of the stau, 
\begin{eqnarray}
\mathcal{L} \simeq \frac{g m_{\tau}}{2 M_W} \mu \tan\beta \tilde{\tau}_L \tilde{\tau}_R^* h^0 + h.c.,
\end{eqnarray}
becomes large and the charge breaking minimum, which can be deeper than the electroweak symmetry breaking minimum, 
might be generated.

The Higgs mass is predicted to be $124-126$\,GeV in the red bands. 
We have taken the SUSY masses for the vector-like matter as $M_{Q'}=M_{U'}=600\, (1200)$\,GeV in the left (right) bands. 
%By considering different values of $M_{Q'}(=M_{U'})$, the Higgs mass can be explained
 %demanding $M_{Q'}=M_{U'} \gtrsim 600$ GeV, the Higgs mass is explained in the region $m_{\tilde{g}} \gtrsim 900$ GeV. 
In the orange (yellow) region, muon $g-2$ is explained at 1 $\sigma$ (2 $\sigma$) level and the corresponding gluino mass is $m_{\tilde{g}} \lesssim 1.3$ (1.8) TeV.  
Since the current results from the SUSY searches exclude the region with $m_{\tilde{g} }\lesssim 1$\,TeV as discussed later, the messenger scale should be lower than $\sim 10^7$ GeV ($\sim 10^{10}$ GeV for 2 $\sigma$) if the SUSY is responsible for the muon $g-2$ anomaly. On the blue dashed line, the lightest nuetralino and the stau are degenerate in the mass. The region above (below) the line, NLSP is the stau (neutralino).

The line $B(M_{\rm mess})=0$ corresponds to the vanishing $B$-term at the messenger scale; the Higgs $B$-term 
 is generated radiatively through the gaugino masses, and the successful electroweak symmetry breaking is achieved for relatively large $\tan\beta$~\cite{Rattazzi:1996fb}. In this case the phases of the B-term and gaugino mass  are aligned, and there is no SUSY CP problem as well as $\mu/B\mu$ problem. 
 
 %%%%%%%%%%%%%%%
%%%%%%%%%%%%%%%

%%%%%%%%%%%%%%%%%%%%%%%%

\begin{figure}
\begin{center}
\includegraphics[width=9cm]{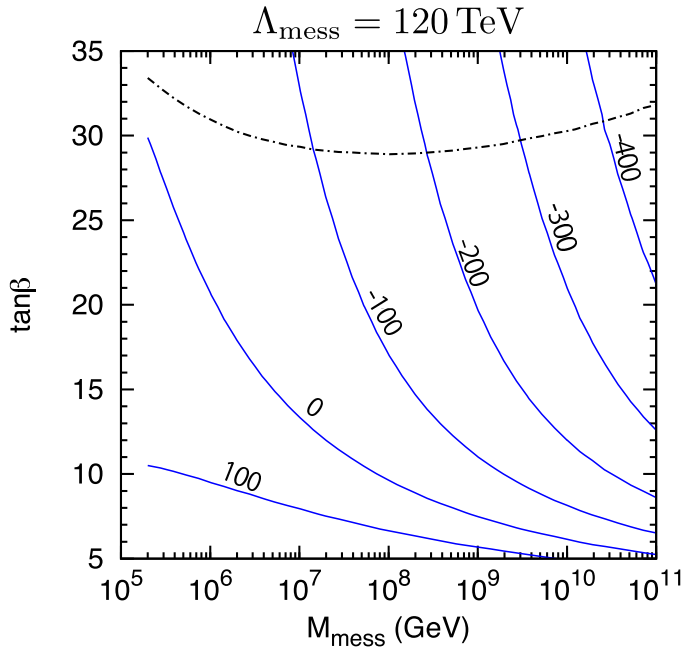}
\caption{The contours of the Higgs $B$-term at the messenger scale, $B({\rm mess})$, is shown in the unit of GeV
on $M_{\rm mess}$-$\tan\beta$ plane. }
\label{fig:mess_B}
\end{center}
\end{figure}

The contours of the Higgs $B$-term at the messenger scale, $B({M_{\rm mess}})$, are shown in Fig.~\ref{fig:mess_B}. As the messenger scale becomes high, the radiative correction from the renormalization group evolution between the messenger scale and the electroweak symmetry breaking scale also becomes large. 
The successful electroweak symmetry breaking requires the following relation at the weak scale
\begin{eqnarray}
\frac{\tan\beta}{1+\tan^2\beta} = \frac{B \mu}{2|\mu|^2 + m_{H_d}^2 + m_{H_u}^2},
\end{eqnarray}
where the radiative corrections are neglected. For large $\tan\beta$, the right hand side of the above equation should be small; the Higgs $B$-term should be small for fixed values of $\mu^2$ and $m_{H_{u,d}}^2$. 
As a result, the low messenger scale is favored, otherwise the generated $B$-term is too large to satisfy the stationary condition for the successful electroweak symmetry breaking.

  %%%%%%%%%%%%%%%
%
%
However even in the case of $B(M_{\rm mess})\neq 0$, the successful electroweak symmetry breaking is realized without generating a dangerous CP violating phase. The additional contribution to the Higgs B-term arises through the following interaction,
\begin{eqnarray}
g_{\bar{D}} {Q}' H_d \Psi_{\bar{D}} + g_{L} \overline{E}' H_d \Psi_{{L}},  \label{eq:hd}
\end{eqnarray}
or 
\begin{eqnarray}
g_{{D}} \overline{Q}' H_u \Psi_{{D}} + g_{\bar{L}} {E}' H_d \Psi_{\bar{L}}, \label{eq:hu}
\end{eqnarray}
where $\Psi_{\bar{D}}$ and $\Psi_{{L}}$ are the parts of the SU(5) multiplet $\overline{\Psi}_{\rm mess}$, transforming ${\bf 3}^* \times {\bf 1}$ and ${\bf 1} \times {\bf 2}$ under $SU(3)_C \times SU(2)_L$, respectively. 
The above interaction (\ref{eq:hd}) ((\ref{eq:hu})) is allowed by choosing the PQ charge of the messenger as $q_{\rm mess}=-2-3q_H/2$ ($q_{\rm mess}=2-3q_H/2$). 
Since the phases of the couplings $g_{\bar{D}}$, $g_{L}$ and $\kappa$ (in Eq.(\ref{eq:pq})) can be removed simultaneously, the above interaction does not generate any CP violating phases. The additional contribution to the Higgs $B$-term is given by~\footnote{
Soft mass squared of $H_{d} (H_u)$ is also modified by the interaction (\ref{eq:hd})((\ref{eq:hu})).
}
\begin{eqnarray}
\delta B(M_{\rm mess}) \simeq - \frac{1}{16\pi^2} (3g_{\bar{D}}^2 + g_{L}^2)\Lambda_{\rm mess},
\end{eqnarray}
with $\Lambda_{\rm mess}=F_X/X$, where we have shown the only leading contribution. 
Since $\delta B$ is negative, the region above the line $B(M_{\rm mess})=0$ in Fig.~\ref{fig:mess_B} is also consistent with the correct electroweak symmetry breaking.

%%%%%%%%%%%%%%%
%%%%%%%%%%%%%%%

%%%%%%%%%%%%%%%%%%%%%%%%%%%%%%%%%%%%%%%%%%%%%%%
%\subsection{SUSY search}

Finally, let us comment on the constraints from the recent results of the SUSY searches. In our model, the muon $g-2$ and the Higgs mass can be explained simultaneously in wide range of the messenger scale, $M_{\rm mess}\simeq {\rm (a~few)} \times 10^{5}$ GeV - $10^{10}$\,GeV. 
For the messenger scale as low as $M_{\rm mess} \sim 10^{5}$\,GeV, the gravitino mass can be as light as $m_{3/2} \sim 10$\,eV
if the dominant SUSY breaking comes from the $F$-term of $X$ ($F_X$).
In this case, the next to the lightest SUSY particle (NLSP) decays into gravitino and SM particle promptly ($c\tau_{\rm NLSP} < 0.1$ mm). 
When the stau is the NLSP, the typical SUSY signal contains large missing transverse momentum, jets and $\tau$ leptons. 
In such a case, the current bound of the gluino mass is obtained as $m_{\tilde{g}} \gtrsim 1$\,TeV~\cite{stau_NLSP}. 
On the other hand, In the region where the bino-like NLSP decaying into the photon and the gravitino, the signal contains large missing transverse momentum and photons, and the constraint is obtained as $m_{\tilde{g}} \gtrsim 1.2$\,TeV for $m_{\tilde{q}} \gtrsim 1.6$\,TeV~\cite{CMS PAS SUSY}. All regions consistent with muon $g-2$ at the 1$\sigma$ level are already excluded.

On the other hand, if the messenger scale is high enough and/or the SUSY breaking $F$-term $(F_{\rm total})$ 
is much larger than $F_X$, the decay length of the NLSP is longer than the detector size. 
Therefore the NLSP can be regarded as the stable particle inside the detector. 
The stau NLSP region which is consistent with muon $g-2$ is expected to be excluded~\cite{Endo:2011xq}.  
When the neutralino is the NLSP, the typical signal contains multi-jets and missing transverse momentum as in the case of mSUGRA models. In such a case, the current bound is estimated as $m_{\tilde{g}} \gtrsim 1$\,TeV for $m_{\tilde{q}} \gtrsim 1.4$\,TeV~\cite{mSUGRA_search} (see also Ref.~\cite{Endo:2011xq}).

\begin{figure}
\begin{center}
\includegraphics[width=9cm]{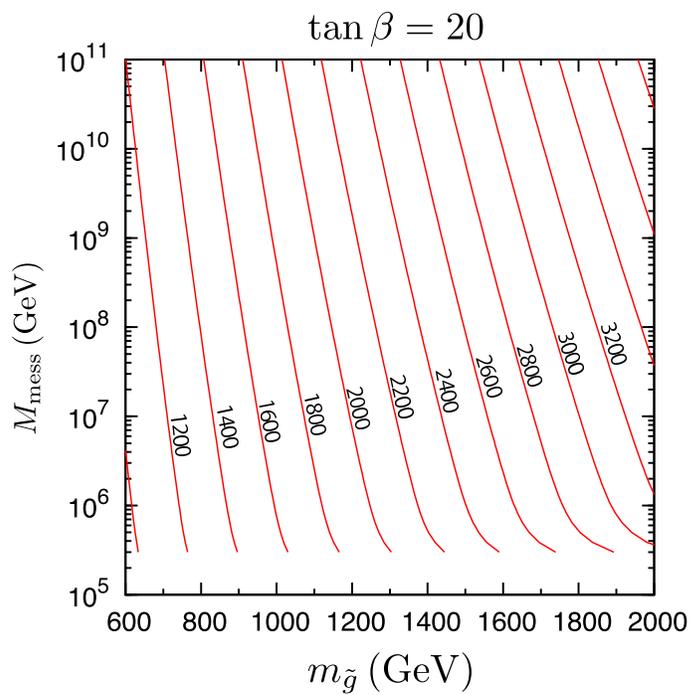}
\caption{The contours of the squark mass on $m_{\tilde{g}}$-$M_{\rm mess}$ plane. The squark masses are shown in the unit of GeV.}
\label{fig:mgl_msq}
\end{center}
\end{figure}

In Fig.~{\ref{fig:mgl_msq}}, contours of the squark masses are shown on $m_{\tilde{g}}-M_{\rm mess}$ plane. 
Current bounds from LHC SUSY search can be avoided for $m_{\tilde{q}} \gtrsim 1$\,TeV and $m_{\tilde{g}} \gtrsim 1.5$\,TeV. 
On the other hand, the muon $g-2$ at the 1$\sigma\, (2\sigma)$ level and the 125\,GeV Higgs boson are explained for $m_{\tilde{g}} \lesssim 1.3 \, (1.8) $\,TeV. 
In our model, the gluino mass $m_{\tilde{g}} \gtrsim 1$\,TeV corresponds to $m_{\tilde{q}} \gtrsim 1.6$\,TeV 
depending on the messenger scale. 
Therefore, the Higgs mass and muon $g-2$ are explained simultaneously, while satisfying the current bound from the LHC SUSY search.

In summary, our model has following properties.
\begin{itemize}

\item The size of $\mu$-term and the masses of vector-like matter are controlled by the PQ symmetry.
They are naturally of the order of TeV for the phenomenologically viable PQ scale of $10^{9}$--$10^{12}$\,GeV, 
hence solves the $\mu$-problem.
The $B$-term can be either zero at the messenger scale or may be generated through the interaction (\ref{eq:hd}) or (\ref{eq:hu}).

\item Due to the radiative correction from additional vector-like matter with masses of $\mathcal O$(TeV), 
the lightest Higgs mass can easily be as heavy as 125\,GeV.
The unwanted coupling of the vector-like matter to the down type Higgs is forbidden by the PQ symmetry.
The muon anomalous magnetic moment can also be explained simultaneously 
while current SUSY searches at the LHC can be avoided.

\end{itemize}

Here are additional comments.

\begin{itemize}

\item The PQ symmetry is anomaly-free for a particle content included in $W_{\rm MSSM}$ and $W_{\rm ext}$.
Thus we have introduced one pair of PQ quarks, $\Psi_{\rm PQ}$ and $\bar\Psi_{\rm PQ}$, 
which transform as ${\bf 5}$ and $\overline {\bf 5}$ under SU(5) in order to make the U(1)$_{\rm PQ}$ anomalous.
This solves the strong CP problem. 
The domain wall number is equal to one and hence we do not suffer from the cosmological domain wall problem.
It also stabilizes the PQ scalar at the scale of $10^{9}$--$10^{12}$\,GeV, as will be seen in the next section.

\item The perturbativity of the gauge couplings is maintained if the mass of heavy quarks, $M_{\rm PQ}=k\langle\phi\rangle$, 
is larger than about $10^{10}$\,GeV for $N_{\rm mess}=1$ and $M_{\rm mess}=10^6$\,GeV. 
Larger $M_{\rm mess}$ leads to the looser constraint on $M_{\rm PQ}$.

%\item The extra matter can decay into SM particles through following operators in the superpotential if $q_H=0$:~\footnote{As another possibility, the decay through the operator, $W=\phi^2/M_{P}^2 (10' \bar{5}_{\rm M} \bar{5}_{\rm M}) $ is also possible if $q_H = -12/5$.
%There are other possible channels depending on the choice of $q_H$.}
%\begin{eqnarray}
%W = \frac{1}{M_{*}} (\bar{Q}' Q + U' \bar{U} + E' \bar{E}) L H_u,
%\end{eqnarray}
%where $M_{*}$ is a cut off scale. For $M_{*} \sim 10^{16}$\,GeV, the lifetimes of the extra matter are $\sim1$\,sec, and it can be consistent with the successful prediction of the big-bang nucleosynthesis (BBN).

\item The seesaw mechanism can work for $q_H=-4/5$. The extra matter can decay into SM particles through higher dimensional operators, with the life time of $\sim 1$ sec, and it can be consistent with the successful prediction of the big-bang nucleosynthesis (BBN) (see Appendix B).

\item The extra-matters can have odd R-parity. In this case, the extra-matters mix with SM particles, and thus they decay into SM particles promptly. (see also recent discussions about discovery of the extra-matters~\cite{Endo:2011xq,ext_search})  

\end{itemize}

%%%%%%%%%%%%%%%%%%%%%%%%%%%%%%%%%%%%

%%%%%%%%%%%%%%%
%%%%%%%%%%%%%%%

%%%%%%%%%%%%%%%%%%%%%%%%%%%%%%%%%%%%%
\section{Stabilization of the PQ scalar and cosmology}  \label{sec:stab}
%%%%%%%%%%%%%%%%%%%%%%%%%%%%%%%%%%%%%

%%%%%%%%%%%%%%%%%%%%%%%%%%%%%%%%%%%%%
\subsection{Stabilizing the PQ scalar}
%%%%%%%%%%%%%%%%%%%%%%%%%%%%%%%%%%%%%

\begin{figure}[tbp]
\begin{center}
\includegraphics[scale=1.6]{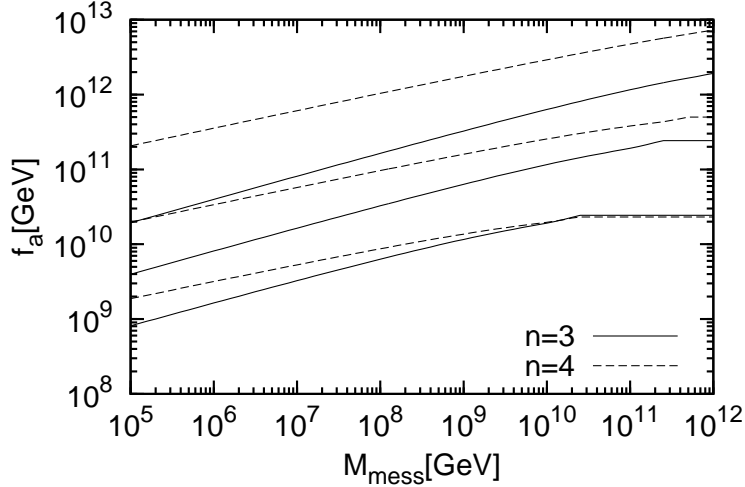}
\caption{
  The PQ scale as a function of $M_{\rm mess}$ for $n=3$ (solid) and $n=4$ (dashed).
  The three lines correspond to $M/M_P=1, 10^2,10^4$ from bottom to top for $n=3$,
  and $M/M_P=10^{-4}, 10^{-2}, 1$ from bottom to top for $n=4$.
}
\label{fig:fa}
\end{center}
\end{figure}

In this section we discuss how to stabilize the PQ scaler at an appropriate scale.
The stabilization of the PQ scalar in the framework of GMSB was discussed e.g. in Refs.~\cite{Asaka:1998ns,ArkaniHamed:1998kj,Banks:2002sd,Carpenter:2009sw,Choi:2011rs,Jeong:2011xu}.
%As shown in the previous section, a relatively low messenger scale 
%$(M_{\rm mess} \lesssim 10^{10}\,{\rm GeV})$ is required in order to realize EWSB.
%Therefore, we assume the PQ scale $\langle \phi\rangle = f_a$ larger than the messenger scale hereafter.
For $k|\phi| \ll M_{\rm mess}(\equiv\kappa X)$, the K\"ahler potential of the PQ scalar below the messenger scale is given by
\begin{equation}
	\mathcal L = \int d^4\theta Z_\phi(X) |\phi|^2,
\end{equation}
where the wave-function renormalization factor $Z_\phi$ depends on the scale $X$ at the three-loop level.\footnote{
	If there is a mixing between the PQ quarks and messenger fields, 
	the correction arises at the one-loop level~\cite{Jeong:2011xu}.
	We do not consider such a case in the following.
}
The dominant contribution to the PQ scalar potential comes from $m_{\phi}^2 (Q=k|\phi| ) |\phi|^2$.
In the opposite case, for $k|\phi| \gg M_{\rm mess}$, the K\"ahler potential of the $X$ field below the PQ scale is given by
\begin{equation}
	\mathcal L = \int d^4\theta Z_X(|\phi|) |X|^2,
\end{equation}
where the wave-function renormalization factor $Z_X$ depends on the scale $|\phi|$ at the three-loop level.
Taking them into account,
the PQ scalar potential is given by~\cite{Asaka:1998ns,ArkaniHamed:1998kj},
\begin{equation}
	|\phi|\frac{\partial V_s}{\partial |\phi|} \simeq \begin{cases} 
		\displaystyle
		-\frac{4k^2}{\pi^2}m_{\rm soft}^2|\phi|^2 \log \left(\frac{M_{\rm mess}}{k|\phi|}\right) & {\rm for}~~~k|\phi| \ll M_{\rm mess} \\
		\displaystyle
		-\frac{g_s^4 \kappa^2}{(4\pi^2)^3} |F_X|^2
		\log^2\left( \frac{k|\phi|}{M_{\rm mess}}\right) & {\rm for}~~~k|\phi| \gg M_{\rm mess} .
	\end{cases}
\end{equation}
where $m_{\rm soft}\equiv(g_s^2/16\pi^2)\Lambda_{\rm mess}$.
This drives the PQ scalar away from the origin.
In order to stabilize the PQ scalar, we introduce the non-renormalizable superpotential
\begin{equation}
	W = \frac{\phi^n \bar\phi}{M^{n-2}},  \label{WNR}
\end{equation}
where $\bar\phi$ has a PQ charge $-n$ with a cutoff scale $M$. 
Then the scalar potential of the PQ sector is given by
\begin{equation}
	V(\phi,\bar\phi) = V_s(|\phi|) + V_{\rm grav} + \frac{ |\phi|^{2(n-1)}}{M^{2(n-2)}}\left(|\phi|^{2} + n^2 |\bar\phi|^2  \right)
	-\left(A_M\frac{\phi^n\bar\phi}{M^{n-2}} + {\rm h.c.}\right),
	\label{scalarpot}
\end{equation}
where $V_{\rm grav} \sim m_{3/2}^2(|\phi|^2 + |\bar\phi|^2)$ represents the gravity-mediation effect
and $A_M \sim {\rm max} \{m_\sigma^2/\Lambda_{\rm mess}$, $m_{3/2}\}$ comes from 
the gauge-mediation effect and the gravity-mediation effect,
where $m_{\sigma}$ denotes the saxion mass defined later.
We choose $A_M$ real and positive by the phase redefinition of the fields.
By minimizing the potential, if the gravity-mediation effect is small enough, we find the PQ scale as
\begin{equation}
	f_a = \langle |\phi| \rangle \simeq
	\begin{cases} 
		\displaystyle \left[ \frac{2k^2}{n\pi^2}m_{\rm soft}^2M^{2n-4}  \log\left( \frac{M_{\rm mess}}{kf_a}\right) \right]^{1/(2n-2)}
		& {\rm if}~~~k f_a \ll M_{\rm mess} \\
		\displaystyle \left[ \frac{\epsilon}{2n}\kappa^2 |F_X|^2 M^{2n-4} \log^2\left(\frac{kf_a}{M_{\rm mess}}\right) \right]^{1/(2n)}
		& {\rm if}~~~k f_a \gg M_{\rm mess} \\
	\end{cases}
\end{equation}
where $\epsilon \equiv g_s^4/(4\pi^2)^3$.
Hereafter we take $k=1$ and $\kappa=1$ for numerical evaluation for simplicity, unless otherwise stated.
The $\bar\phi$ field also obtains a VEV due to the $A$-term, as
\begin{equation}
	\bar v \equiv\langle|\bar\phi|\rangle = \frac{A_M M^{n-2}}{n^2f_a^{n-2}}.   \label{barphi}
\end{equation}
Notice that this should be much smaller than $f_a$ in order for the calculation so far to remain valid.
Actually, $\bar v / f_a \sim A_M/m_\sigma \ll 1$ is always satisfied.
If the $X$ dominantly breaks the SUSY, we have a relation
\begin{equation}
	f_a \simeq 6\times 10^{9}\,{\rm GeV}\left( \frac{m_{3/2}}{1\,{\rm keV} }\right)^{1/3}
	\left( \frac{M}{M_P} \right)^{1/3},
	\label{fa_n3}
\end{equation}
for $n=3$ and
\begin{equation}
	f_a \simeq 9\times 10^{11}\,{\rm GeV}\left( \frac{m_{3/2}}{1\,{\rm keV} }\right)^{1/4}
	\left( \frac{M}{M_P} \right)^{1/2},
	\label{fa_n4}
\end{equation}
for $n=4$, if $k f_a \gg M_{\rm mess}$.
Thus we can have a correct value of the PQ scale for $M\sim M_P$.
If the dominant SUSY breaking is carried by another field, we can obtain a correct PQ scale for larger gravitino mass.
In the large gravitino mass limit, $V_{\rm grav}$ tends to dominate the potential and it determines the PQ scale.
Fig.~\ref{fig:fa} shows the PQ scale $f_a$ as a function of $M_{\rm mess}$
for $n=3$ (solid) and $n=4$ (dashed).
The three lines correspond to $M/M_P=1, 10^2,10^4$ from bottom to top for $n=3$, 
and $M/M_P=10^{-4}, 10^{-2}, 1$ from bottom to top for $n=4$. 
We have taken $\Lambda_{\rm mess}=100$\,TeV and $F_X = \sqrt{3}m_{3/2}M_P$.
%Since the $B$-term is small at the messenger scale, we need $M_{\rm mess}\lesssim 10^8$\,GeV as described in the
%previous section. Taking uncertainties of the parameters such as $M, k, \kappa$ into account,  
%we roughly have a PQ scale as $10^9\,{\rm GeV}\lesssim f_a \lesssim 10^{10}$\,GeV for $n=3$ and 
%$10^{10}\,{\rm GeV}\lesssim f_a \lesssim 10^{11}$\,GeV for $n=4$.
%This constraint is relaxed if we allow some mechanism to generate the $B$-term at the messenger scale.

The angular component of the $\phi$ around the minimum $f_a$ is regarded as the QCD axion,
which dynamically solves the strong CP problem.\footnote{
	To be more precise, it is a linear combination of the angular components of $\phi$ and $\bar\phi$ that is regarded as the axion.
	In the limit $\langle\phi\rangle \gg \langle\bar\phi\rangle$, however, it mostly consists of the angular component of $\phi$.
}
On the other hand, the fluctuation in the radial direction, which we denote by $\sigma$,
is called the saxion.
In the present setup, the saxion mass is given by
\begin{equation}
	m_\sigma \simeq \sqrt{2}n\frac{f_a^{n-1}}{M^{n-2}}.
\end{equation}
It is evaluated as
\begin{equation}
	m_\sigma \simeq 174\,{\rm GeV} 
	\left( \frac{f_a}{10^{10}\,{\rm GeV}} \right)^2 \left( \frac{M_P}{M} \right),
\end{equation}
for $n=3$ and
\begin{equation}
	m_\sigma \simeq 0.95 \,{\rm GeV} 
	\left( \frac{f_a}{10^{12}\,{\rm GeV}} \right)^3 \left( \frac{M_P}{M} \right)^2,
\end{equation}
for $n=4$.
If the $X$ dominantly breaks SUSY, we obtain the saxion mass by using (\ref{fa_n3}) and (\ref{fa_n4}) as
\begin{equation}
	m_\sigma \simeq 60\,{\rm GeV} 
	\left( \frac{m_{3/2}}{1\,{\rm keV} }\right)^{2/3}
	\left( \frac{M_P}{M} \right)^{1/3},
\end{equation}
for $n=3$ and
\begin{equation}
	m_\sigma \simeq 0.7\,{\rm GeV} 
	\left( \frac{m_{3/2}}{1\,{\rm keV} }\right)^{3/4}
	\left( \frac{M_P}{M} \right)^{1/2},
\end{equation}
for $n=4$, if $kf_a \gg M_{\rm mess}$.
The axino, a fermonic superpartner of the axion, obtains a mass from the operator (\ref{WNR}).
Substituting (\ref{barphi}) back into (\ref{WNR}), we obtain the axino mass of 
$m_{\tilde a}\simeq m_{\sigma}/\sqrt{2}$, taking into account the mixing of $\tilde \phi$ and $\tilde{\bar\phi}$.\footnote{
	The $\tilde \phi$ and $\tilde{\bar\phi}$ mostly mix with eath other in the mass eigenstate.
	We call them as the axino and denote them by $\tilde a$.
}
The axino also obtains a mass radiatively at the three-loop level, 
but the contribution is suppressed by the three-loop factor and the ratio $M_{\rm mess}/f_a$
compared with the gaugino mass, which is much smaller than the tree-level mass.

%%%%%%%%%%%%%%%
%%%%%%%%%%%%%%%

%%%%%%%%%%%%%%%%%%%%%%%%%%%%%%%%%%%%%
\subsection{Cosmology}
%%%%%%%%%%%%%%%%%%%%%%%%%%%%%%%%%%%%%

Let us discuss the cosmological implications of the present model.
In particular, we focus on the behavior of saxion.
The saxion decays into the axion pair, gauge boson pair, SM fermion pair and the Higgs boson pair
as long as they are kinematically allowed.
The decay rates of saxion and $\bar\phi$ are summarized in the Appendix.
%For most parameter regions we are interested in ($10^{10}$\,GeV$\lesssim f_a\lesssim 10^{12}$\,GeV
%and $1$\,GeV$\lesssim m_\sigma\lesssim 100$\,GeV), the saxion decays before BBN begins.
Since we are interested in the saxion mass range $m_\sigma \lesssim 100$\,GeV,
the decay into the SUSY particle pair, including the axino pair, is kinematically forbidden.
Let us discuss the cosmology in two cases : (i) the saxion is initially trapped at the origin, and
(ii) the saxion is initially far from the origin.
\\

\subsubsection{The saxion trapped at the origin}

First let us suppose that the saxion obtains a positive Hubble mass squared in the early Universe:
$V\sim H^2 |\phi|^2$.
It sits at the origin during and after inflation, and the PQ quarks are massless there.
Therefore, PQ quarks are thermalized and give the thermal mass to the PQ scalar $\phi$.
The situation lasts until the temperature drops to the weak scale,
where the instability of the PQ scalar develops and the PQ symmetry is broken.
In such a case, the potential energy at the origin dominates the Universe
before the PQ phase transition, which leads to a short-lasted period of inflation : thermal inflation~\cite{Yamamoto:1985rd,Lyth:1995hj}.
(Thermal inflation in the context of PQ symmetry breaking was discussed in Refs.~\cite{Asaka:1998xa,Chun:2000jr,Kim:2008yu,Choi:2009qd,Park:2010qd}.)
All the dangerous relics such as the gravitino and axino are diluted away.
The Universe is reheated by the decay of the saxion.
For successful reheating, the saxion must decay dominantly into fermions or Higgs bosons
before BBN begins.
Otherwise, decay-produced axions contribute too much to the effective number of neutrino species.
For example, the saxion may dominantly decay into $\tau\bar\tau$ pair for $m_\sigma \sim 4$\,GeV, 
and into $b\bar b$ pair for $m_\sigma \sim 9$\,GeV.
The successful reheating is achieved for such mass ranges.
Note that there are no axionic domain walls after the phase transition since in this model 
the domain wall number is equal to one.
According to the most recent estimate~\cite{Hiramatsu:2012gg},
the PQ scale is restricted as $f_a \lesssim 2\times 10^{10}$\,GeV taking account of the 
abundance of the axion emitted from the axionic strings and collapsing domain walls if the saxion decays before the QCD phase transition.
If the saxion decays after the QCD phase transition, the axion is diluted by the saxion decay and the 
constraint is relaxed~\cite{Kawasaki:1995vt}.

The $\bar\phi$ field also begins to oscillate around its minimum with an amplitude of $\bar v$
after thermal inflation ends. Although it has longer lifetime than the saxion by the factor $\sim (f_a/\bar v)^2$ (see Appendix),
the abundance is suppressed by the factor $\sim (\bar v/f_a)^2$.
As a result, $\bar\phi$ decays before it comes to dominate the Universe.
Therefore, it does not drastically change the picture as long as $\bar\phi$ decays before BBN.
Otherwise, the $\bar\phi$ decay can have problematic effects on BBN.

One should also care about the existence of heavy stable PQ matter, $\Psi_{\rm PQ}$, 
which once were in thermal equilibrium during thermal inflation, supporting the PQ scalar at the origin.
In particular, the mass of the neutral component in $\Psi_{\rm PQ}$ is severely constrained
in order for it not to be overabundant~\cite{Moroi:2011be}.

Baryon number is also diluted away by thermal inflation.
A variant type of the Affleck-Dine mechanism may work for generating the baryon asymmetry again after thermal inflation~\cite{Stewart:1996ai,Jeong:2004hy,Felder:2007iz,Kim:2008yu}.\\

\subsubsection{The saxion far from the origin}

Next we consider the case where the saxion is displaced far from the origin
because of the negative Hubble correction : $V\sim -H^2 |\phi|^2$.
Then the PQ scalar tracks the minimum $|\phi|\sim (HM^{n-2})^{1/(n-1)}$ as the Hubble parameter decreases after inflation.
In this case the saxion relaxes to the true vacuum while the PQ symmetry is never restored.
The saxion oscillation is induced at $H\sim m_\sigma$, with an amplitude of $\sim f_a$ around the minimum.
The saxion abundance is then given by
 \begin{equation}
 \begin{split}
	\frac{\rho_\sigma}{s} &\simeq \frac{\epsilon}{8}T_{\rm R}\left( \frac{f_a}{M_P} \right)^2
	\simeq 2\times 10^{-11}\,{\rm GeV}
	\left( \frac{T_{\rm R}}{10^5\,{\rm GeV}} \right)
	\left( \frac{f_a}{10^{11}\,{\rm GeV}} \right)^2\epsilon,
\end{split}
\end{equation}
where $\epsilon =1$ for $T_{\rm R} \lesssim \sqrt{m_\sigma M_P}$ and  
$\epsilon \simeq\sqrt{m_\sigma M_P}/T_{\rm R}$ for $T_{\rm R} \gtrsim \sqrt{m_\sigma M_P}$.
Depending on the saxion lifetime and the branching ratio into visible particles, 
we obtain an upper bound on the reheating temperature $T_{\rm R}$~\cite{Kawasaki:2007mk}.
The saxion decay temperature, $T_\sigma$, is given by
\begin{equation}
	T_\sigma = \left( \frac{10}{\pi^2 g_*(T_{\sigma})} \right)^{1/4}\sqrt{\Gamma_\sigma M_P}
	\sim \begin{cases}
		\displaystyle
		3\,{\rm GeV} \left( \frac{m_{3/2}}{1\,{\rm keV} }\right)^{2/3}
		\left( \frac{M_P}{M} \right)^{5/6}& {\rm for }~n=3 \\
		\displaystyle
		3\times 10^{-5}\,{\rm GeV} \left( \frac{m_{3/2}}{1\,{\rm keV} }\right)^{7/8}
		\left( \frac{M_P}{M} \right)^{5/4}& {\rm for }~n=4	
	\end{cases}.
\end{equation}
where $\Gamma_\sigma$ is the total decay width of the saxion.
In this expression we have assumed that the saxion dominantly decays into axion pair, 
and also substituted (\ref{fa_n3}) and (\ref{fa_n4}).
If the saxion decays before BBN, the bound comes from 
the requirement that the effective number of neutrino species, $\Delta N_{\rm eff}$, must not be much larger than one. However, it does not pose a severe constraint.
On the other hand, if the saxion decays after BBN, it may affect the primordial light element abundances
through its hadronic or radiative decay processes
and hence the saxion abundance is constrained~\cite{Kawasaki:2004qu}.

Besides saxion, the $\bar\phi$ coherent oscillation is also induced.
As the VEV of $\phi$ decreases, the $\bar\phi$ tracks the temporal minimum determined by
$\bar\phi~(\sim A_M M^{n-2}/|\phi|^{n-2})$.
At $H\sim m_\sigma$ where the $\phi$ begins to oscillate around the minimum,
the $\bar\phi$ also begins to oscillate around the true minimum.
The typical oscillation amplitude is estimated to be $\bar v$ given in (\ref{barphi}),
which is much smaller than $f_a$.
Although the abundance of $\bar\phi$ is much smaller than the saxion coherent oscillation,
its lifetime is much longer, as shown in Appendix, and hence it is nontrivial whether $\bar\phi$ poses a severe constraint.
Actually, as will be shown below, $\bar\phi$ can give severer constraint than the saxion depending on their masses.

The axino $(\tilde a)$ is produced by scattering of particles in thermal bath.
Since we have PQ quarks that couples to the PQ scalar, axinos are produced through the 
axino-gluon-gluino interaction during the reheating~\cite{Covi:2001nw}.
The axino abundance produced by the gluon scattering is proportional to the reheating temperature, and given by
\begin{equation}
	Y_{\tilde a}^{(g)}\simeq 2\times 10^{-6} g_s^6
	%\ln \left( \frac{1.108}{g_s} \right)
	\left( \frac{f_a}{10^{11}\,{\rm GeV}} \right)^{-2}
	\left( \frac{T_{\rm R}}{10^{5}\,{\rm GeV}} \right).
	\label{Yag}
\end{equation}
In addition, the axino has a tree-level coupling to the Higgs fields and vector-like matter.
The scattering of Higgs fields and vector-like matter produce axinos as long as the temperature is higher than 
the masses of the Higgs/higgsino and vector-like matter.
This contribution is roughly given by~\cite{Chun:2011zd,Bae:2011jb}
\begin{equation}
	Y_{\tilde a}^{(h)}\simeq 10^{-5}
	%\ln \left( \frac{1.108}{g_s} \right)
	\left( \frac{f_a}{10^{11}\,{\rm GeV}} \right)^{-2}
	\left( \frac{m_{\rm vec}}{1\,{\rm TeV}} \right)^2,
	\label{Yah}
\end{equation}
where $m_{\rm vec}$ represents the higgsino mass ($\mu$) or the mass of vector-like particles, whichever is heavier.
The axino decays into the gravitino, with the rate
\begin{equation}
	\Gamma(\tilde a \to a \psi_{3/2}) = \frac{1}{96\pi}\frac{m_{\tilde a}^5}{m_{3/2}^2M_P^2}
	\simeq (1.1\times 10^3\,{\rm sec})^{-1}
	\left( \frac{m_{\tilde a}}{1\,{\rm GeV}} \right)^{5}
	\left( \frac{1\,{\rm keV}}{m_{3/2}} \right)^{2}.
\end{equation}
Since the axino decays into the gravitino and axion, the bound reads
$m_{3/2}Y_{\tilde a}\lesssim 4\times 10^{-10}\,{\rm GeV}$.\footnote{
	The gravitino produced by the axino decay has a long free-streaming length
	and behave as warm/hot dark matter rather than the cold dark matter, depending on their masses.
	In this case the constraint is severer.
}
For relatively small axino mass and large gravitino mass, the axino lifetime becomes so long that
it dominates the Universe before it decays.
Thus, the constraint that axions produced by the axino decay do not contribute too much to the $N_{\rm eff}$
also gives upper bound on the reheating temperature.
%These contributions are too large for $f_a \lesssim 10^{10}$\,GeV except for the case of ultra-light gravitino 
%$(m_{3/2}\sim 10\,{\rm eV})$, unless the reheating temperature is so small that SUSY particles never reach thermal equilibrium.
%For $f_a\sim 10^{11}$\,GeV, the gravitino mass of $m_{3/2} \sim 1$\,keV is allowed.
%However, taking account of the gravitino thermal production,
%we need a low-reheating temperature $T_{\rm R} \lesssim 1$\,TeV in order to avoid the gravitino overproduction.
%At the boundary of the constraint, the gravitino can account for DM.

Note also that a similar process results in the saxion thermal production and the
saxion abundance is comparable to the estimates (\ref{Yag}) and (\ref{Yah}).
Moreover, $\bar\phi$ particles are also produced similarly, whose abundance is suppressed by the 
factor $\sim (\bar v/f_a)^2$.
Although the abundance of $\bar\phi$ is much smaller than the saxion and axino,
its lifetime is much longer and hence it is nontrivial whether thermally produced $\bar\phi$ poses a severe constraint.

We derive cosmological constraints on the present model taking into account all the above mentioned contributions :
the coherent oscillation of saxion and $\bar\phi$, thermally produced saxion, $\bar\phi$ and axino.
We follow the methods in Ref.~\cite{Kawasaki:2007mk} to derive these constraints,
using decay rates calculated in Appendix.
The upper bound on $T_{\rm R}$ is obtained from the requirement that their decay products 
do not contribute to $\Delta N_{\rm eff}$ and DM abundance too much, do not disturb BBN, 
do not distort the blackbody spectrum of cosmic microwave background,
do not yield too much X($\gamma$)-ray background.
The bound from gravitino overproduction~\cite{Bolz:2000fu} is also considered.
See Ref.~\cite{Kawasaki:2007mk} for details.

\begin{figure}[tbp]
\begin{center}
\includegraphics[scale=1.6]{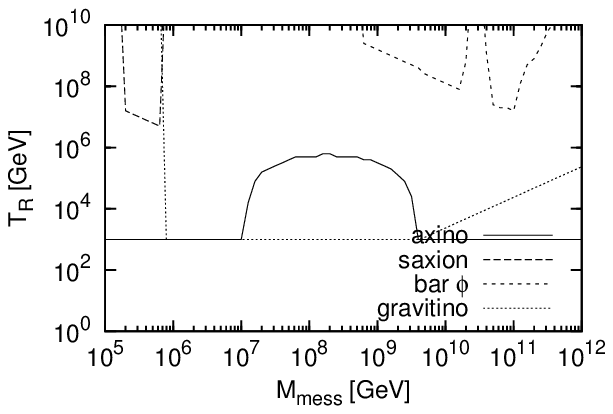}
\includegraphics[scale=1.6]{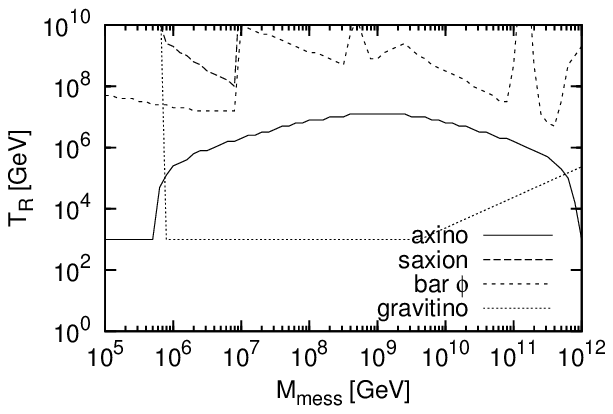}
\caption{
	Constraints on the reheating temperature $T_{\rm R}$ as a function of $M_{\rm mess}$
	for $n=3$ and $M=10^4M_P$ (top), and $n=4$ and $M=M_P$ (bottom).
	Each line corresponds to the bound from the axino, saxion, $\bar\phi$ and gravitino.
}
\label{fig:TR}
\end{center}
\end{figure}

Fig.~\ref{fig:TR} shows constraints on the reheating temperature $T_{\rm R}$ as a function of $M_{\rm mess}$
for $n=3$ and $M=10^4M_P$ (top) and $n=4$ and $M=M_P$ (bottom).
Each line corresponds to the bound from the axino, saxion, $\bar\phi$ and gravitino.
We have also taken $F_X = \sqrt{3}m_{3/2}M_P$.
A characteristic behavior of the axino bound comes from the fact that, for lower messenger scale, the axino becomes light and its lifetime is too long and hence it tends to dominate the Universe. 
For higher messenger scale, the gravitino becomes heavy
and its abundance coming from the axino decay tends to be too large.
Therefore, the bound from axino is relatively weak at intermediate messenger scale.
The bounds from saxion and $\bar\phi$ shows a complicated behavior since their lifetimes
significantly change at the threshold for the decay into quark/leptons and also the BBN and other constraints 
significantly depend on the lifetime of decaying particles.
In the most parameter space, the reheating temperature is bounded as $T_{\rm R}\lesssim 1$\,TeV
in order to avoid the axino and gravitino overproduction.
Constraints from the saxion and $\bar\phi$ are less stringent.

%%%%%%%%%%%%%%%
%%%%%%%%%%%%%%%

Finally, we mention a constraint on the inflation model.
The axion obtains an isocurvature fluctuation during inflation in this case.
The magnitude of the CDM isocurvature perturbation is given by
\begin{equation}
	|S_{\rm c}|  = \frac{\Omega_a}{\Omega_{\rm c}}\frac{2\delta\theta}{\theta} \simeq
	 \frac{\Omega_a}{\Omega_{\rm c}\theta} \frac{H_{\rm inf}^{1/2}}{\pi M^{1/2}}
	 \simeq 2\times 10^{-6} \frac{\Omega_a}{\Omega_{\rm c}\theta}
	 \left( \frac{H_{\rm inf}}{10^{8}\,{\rm GeV}} \right)^{1/2}
	 \left( \frac{M_P}{M} \right)^{1/2}.
\end{equation}
for $n=3$ and
\begin{equation}
	|S_{\rm c}|  \simeq
	 \frac{\Omega_a}{\Omega_{\rm c}\theta} \frac{H_{\rm inf}^{2/3}}{\pi M^{2/3}}
	 \simeq 4\times 10^{-8} \frac{\Omega_a}{\Omega_{\rm c}\theta}
	 \left( \frac{H_{\rm inf}}{10^{8}\,{\rm GeV}} \right)^{2/3}
	 \left( \frac{M_P}{M} \right)^{2/3}.
\end{equation}
for $n=4$,
where $H_{\rm inf}$ is the Hubble scale during inflation, $\Omega_a$ and $\Omega_{\rm c}$
denote the density parameters of the axion and CDM, respectively, $\theta$ the initial misalignment angle.
The axion abundance is given by~\cite{Preskill:1982cy,Turner:1985si,Bae:2008ue}
\begin{equation}
	\frac{\Omega_a}{\Omega_{\rm c}} \simeq 
	 \left( \frac{f_{a}}{10^{11}\,{\rm GeV}} \right)^{1.18}\theta^2.
\end{equation}
Comparing it with the bound $|S_{\rm c}|<1.4\times 10^{-5}$ from the WMAP 7 year result  
combined with the baryon acoustic oscillation and the measurement of the Hubble constant~\cite{Komatsu:2010fb}, 
we obtain an upper bound on the inflation scale.
If the axion is the dominant component of DM, it may be close to the current upper bound.\footnote{
	If the misalignment angle $\theta$ is small while keeping the magnitude of $S_{\rm c}$, there can be large non-Gaussianity 
	in the CDM isocurvature perturbation~\cite{Kawasaki:2008sn}.
}

%%%%%%%%%%%%%%%%%%%%%%%%%%%%%%%%%%%%%
\section{Conclusions and discussion}  \label{sec:conc}
%%%%%%%%%%%%%%%%%%%%%%%%%%%%%%%%%%%%%

In this paper we have presented a Peccei-Quinn extended GMSB model in which the extra vector-like matters are introduced in order to explain the 125\,GeV Higgs boson.
The PQ symmetry controls the size of $\mu$-term as well as the masses of the vector-like matters.
These parameters have correct size for the PQ symmetry breaking scale of $10^{10}$--$10^{12}$\,GeV.
Thus the strong CP problem as well as the SUSY CP problem are solved in our model.
Fortunately, the PQ symmetry forbids the unwanted coupling between the vector-like matter and the down type Higgs,
which would otherwise make negative contribution to the Higgs boson mass and induce a CP violating phase.
There are parameter regions where the muon $g-2$ anomaly is explained by the SUSY contribution
consistently with the 125\,GeV Higgs boson.
We have also constructed a method to stabilize the PQ scalar and shown that 
the correct PQ scale can be obtained for natural parameter choices.
We have discussed cosmological effects of the saxion and axino, and derived 
upper bound on the reheating temperature. 

Some notes are in order.
Although we have focused on the GMSB model, a similar model can also be applied to the 
gravity-mediation models such as the CMSSM.
The Higgs mass as well as the muon $g-2$ can be explained simultaneously.
Although there are no messenger fields, the stabilization of the PQ scalar is achieved by the
balance between the negative gravity-mediated mass term $V \sim -m_{3/2}^2|\phi|^2$
and the non-renormalizable potential arising from Eq.~(\ref{WNR}).
As another method, we may simply introduce the superpotential as
\begin{equation}
	W = \kappa' S (\phi\bar\phi-f_a^2),
\end{equation}
where $S$ is a singlet field. In this case, the positive gravity-mediated mass term $V \sim m_{3/2}^2(|\phi|^2+|\bar\phi|^2)$
stabilizes the PQ scalars at $|\phi|\sim|\bar\phi|\sim f_a$.
This may also be consistent with the hybrid inflation model of Ref.~\cite{Kawasaki:2010gv}.

%%%%%%%%%%%%%%%%%%%%%%%%%%%%%%%%%%%%
\section*{Acknowledgment}
%%%%%%%%%%%%%%%%%%%%%%%%%%%%%%%%%%%%
This work is supported by Grant-in-Aid for Scientific research
from the Ministry of Education, Science, Sports, and Culture (MEXT),
Japan, No. 24-7523 [NY], No. 21111006 [KN] and No. 22244030 [KN].

%%%%%%%%%%%%%%%%%%%%%%%%%%%%%%%%%%%%

%%%%%%%%%%%%%%%%%%%%%%%%%%%%%%%%%%%%
\appendix
%%%%%%%%%%%%%%%%%%%%%%%%%%%%%%%%%%%%

%%%%%%%%%%%%%%%%%%%%%%%%%%%%%%%%%%%%
\section{Decay of saxion and $\bar\phi$}
%%%%%%%%%%%%%%%%%%%%%%%%%%%%%%%%%%%%

In this appendix we list up the decay modes of the saxion and $\bar\phi$.
First, we consider the mixing between PQ scalars, $\phi$ and $\bar\phi$.
Let us expand them as
\begin{eqnarray}
	\phi        &=& f_a +\frac{1}{\sqrt{2}}(\sigma+ia), \\
	\bar\phi &=& \bar v+\frac{1}{\sqrt{2}}(\bar\sigma+i\bar a).
\end{eqnarray}
They couple in the superpotenial as $W = \phi^n \bar\phi / M^{n-2}$.
By noting the relation (\ref{barphi}), the scalar potential (\ref{scalarpot}) is expanded as
\begin{eqnarray}
	V = \frac{1}{2}\vec \sigma^t \mathcal M_\sigma^2 \vec \sigma+\frac{1}{2}\vec a^t \mathcal M_a^2 \vec a
	+\frac{m_\sigma^2}{2\sqrt{2} f_a}\sigma a^2+\frac{m_{\bar\sigma}^2}{2\sqrt{2} \bar v}\bar\sigma \bar a^2 
	+\frac{(n-1)A_M m_{\bar\sigma}}{\sqrt{2}n}\left( \frac{1}{f_a}\bar \sigma a^2 + \frac{1}{\bar v} \sigma \bar a^2 \right),
	\label{lag}
\end{eqnarray}
where $\vec \sigma=(\sigma,\bar\sigma)$, $\vec a =(a, \bar a)$ and
\begin{eqnarray}
	\mathcal M_\sigma^2 
		&=& \begin{pmatrix} 
		m_\sigma^2 & \frac{n-2}{n}A_M m_{\bar\sigma} \\
		\frac{n-2}{n}A_M m_{\bar\sigma}  & m_{\bar\sigma}^2
	\end{pmatrix}, \\
	\mathcal M_a^2 
		&=& \begin{pmatrix} 
		n A_M m_{\bar\sigma}(\bar v/f_a)  & A_M m_{\bar\sigma} \\
		A_M m_{\bar\sigma}  &  (1/n)A_M m_{\bar\sigma} (f_a/\bar v)
	\end{pmatrix},
\end{eqnarray}
where 
\begin{equation}
\begin{split}
	m_\sigma^2 &= \frac{1}{2}\left.\frac{\partial^2 V_s(|\phi|)}{\partial |\phi|^2}\right|_{|\phi|=f_a}
	+\frac{\partial^2 V_{\rm grav}(|\phi|)}{\partial |\phi|^2}
	+n(2n-1)\frac{f_a^{2(n-1)}}{M^{2(n-2)}},\\
	m_{\bar\sigma}^2 &= n^2\frac{f_a^{2(n-1)}}{M^{2(n-2)}}.
\end{split}
\end{equation}
As for the CP-odd parts, there is a massless mode since ${\rm det} (\mathcal M_a^2) = 0$, as is expected from the PQ symmetry. In what follows, we approximate as $m_\sigma \simeq \sqrt{2}nf_a^{n-1}/M^{n-2}$. Then we obtain
\begin{eqnarray}
	\mathcal M_\sigma^2 
		&\simeq& \begin{pmatrix} 
		m_\sigma^2 & \frac{n-2}{\sqrt{2}n}A_M m_\sigma \\
		\frac{n-2}{\sqrt{2}n}A_M m_\sigma & m_\sigma^2/2
	\end{pmatrix}, \\
	\mathcal M_a^2 
		&\simeq& \begin{pmatrix} 
		\frac{n^2}{2}m_\sigma^2 \frac{\bar v^2}{f_a^2} & \frac{1}{\sqrt{2}}A_M m_\sigma \\
		\frac{1}{\sqrt{2}}A_M m_\sigma & m_\sigma^2/2
	\end{pmatrix}.
\end{eqnarray}

For $f_a \gg \bar v$, the mass eigenstates are given by
\begin{eqnarray}
	\tilde a        &\simeq& a -\frac{n\bar v}{f_a}\bar a, \\
	\tilde{\bar a} &\simeq& \bar a + \frac{n \bar v}{f_a} a,
	\label{a-abarmix}
\end{eqnarray}
where $\tilde a$ is the massless Goldstone mode, which is regarded as the axion,
while $\tilde{\bar a}$ has a mass of $m_\sigma/\sqrt{2}$.
In the most part of this paper, we have not distinguished $a$ and $\tilde a$ since the mixing angle is small.

As for the CP-even part, the mass eigenstates read
\begin{eqnarray}
	\tilde\sigma        &\simeq &  \sigma + \frac{\sqrt{2}n(n-2)A_M}{m_\sigma}\bar\sigma
	=   \sigma + \frac{n^2(n-2)\bar v}{f_a}\bar\sigma, \\
	\tilde{\bar\sigma} &\simeq &   \bar\sigma - \frac{\sqrt{2}n(n-2)A_M}{m_\sigma}\sigma
	= \bar\sigma - \frac{n^2(n-2)\bar v}{f_a}\sigma,
	\label{s-sbarmix}
\end{eqnarray}
where the saxion $\tilde\sigma$, which mostly consists of $\sigma$, has a mass of $m_\sigma$
and $\tilde{\bar \sigma}$ has a mass of $m_{\sigma}/\sqrt{2}$.

%%%%%%%%%%%%%%%%%%%%%%%%%%%%%%%%%%%%
\subsection{Saxion decay}
%%%%%%%%%%%%%%%%%%%%%%%%%%%%%%%%%%%%

Here we summarize the saxion decay mode.
Since the saxion is at most around $\mathcal O(100)$\,GeV in our model,
we neglect the decay into SUSY particles.
We also ignore the mixing of $\sigma$ and $\bar\sigma$ in this subsection,
since it does not affect the saxion decay rate summarized in the following as long as the mixing angle is small.

%%%%%%%%%%%%%%%%%%%%%%%%%%%%%%%%%%%%
\subsubsection{Decay into axions}
%%%%%%%%%%%%%%%%%%%%%%%%%%%%%%%%%%%%

From the Lagrangian (\ref{lag}), the saxion decay rate into axions is calculated as\footnote{
	If the PQ scalar is expanded as $\phi = f_a \exp\left[(\sigma+ia)/\sqrt{2}\right]$, 
	we find the same decay rate from the kinetic term $\mathcal L = |\partial \phi|^2$.
}
\begin{equation}
	\Gamma(\sigma\to 2a)=\frac{1}{64\pi}\frac{m_\sigma^3}{f_a^2}.
\end{equation}
Numerically, it is evaluated as
\begin{equation}
	\Gamma(\sigma\to 2a)
	\simeq (1\times 10^{-2}\,{\rm sec})^{-1}
	\left( \frac{m_\sigma}{1\,{\rm GeV}} \right)^3
	\left( \frac{10^{10}\,{\rm GeV}}{f_a} \right)^2.
	\label{lifetime}
\end{equation}
%%

%%%%%%%%%%%%%%%%%%%%%%%%%%%%%%%%%%%%
\subsubsection{Decay into gluons}
%%%%%%%%%%%%%%%%%%%%%%%%%%%%%%%%%%%%

if the saxion is heavier than $\sim 1$\,GeV, it can decay into the gluon pair, which hadronize
and results in the production of energetic particles.
Integrating out the PQ quarks yields the following couplings
\begin{equation}
	\mathcal L= \frac{\alpha_s}{8\pi}\left( \frac{\sigma}{f_a}G_{\mu\nu}^aG^{\mu\nu a}
	+\frac{a}{f_a}G_{\mu\nu}^a\tilde G^{\mu\nu a}\right).
	\label{aGG}
\end{equation}
This induces the decay into gluons as
\begin{equation}
	\Gamma(\sigma\to 2g)\simeq\frac{\alpha_s^2}{32\pi^3}\frac{m_\sigma^3}{f_a^2}.
\end{equation}
%%

%%%%%%%%%%%%%%%%%%%%%%%%%%%%%%%%%%%%
\subsubsection{Decay into fermions}
%%%%%%%%%%%%%%%%%%%%%%%%%%%%%%%%%%%%

Let us expand the Higgs and PQ scalar as
\begin{eqnarray}
	H_u^0 &=& v_u +\frac{1}{\sqrt{2}}(h_u+ia_u), \\
	H_d^0 &=& v_d +\frac{1}{\sqrt{2}}(h_d+ia_d).
\end{eqnarray}
They have a coupling in the superpotential as
\begin{equation}
	W = \frac{\lambda \phi^\ell}{M_P^{\ell-1}}H_u H_d.
\end{equation}
We have focused on the case of $\ell=2$ in this paper, but here we do not fix it.
The $\mu$-term is generated through the VEV of $\phi$ as $\mu = \lambda f_a^\ell / M_P^{\ell-1}$.
Let us focus on the CP-even parts.
First, by diagonalizing the $h_u$ and $h_d$, we obtain the light and heavy Higgs bosons as
\begin{eqnarray}
	h = h_u \cos \alpha - h_d\sin\alpha, \\
	H = h_u \sin \alpha + h_d\cos\alpha.
\end{eqnarray}
They mix with the saxion $\sigma$ in the mass eigenstates.
Writing the mass eigenstates as $\tilde h$ and $\tilde H$, we find
\begin{eqnarray}
	\tilde h &\simeq&  h -\frac{\ell\mu v }{f_a (m_h^2-m_\sigma^2)}\left[
		B \cos(\alpha+\beta)+2\mu\sin(\alpha-\beta) \right] \sigma, \\
	\tilde H&\simeq& H +\frac{\ell\mu v }{f_a (m_H^2-m_\sigma^2)}\left[
		-B \sin(\alpha+\beta)+2\mu\cos(\alpha-\beta) \right] \sigma,
\end{eqnarray}
where $v\equiv (v_u^2+v_d^2)^{1/2}$.
From these mixings, we can calculate the saxion decay rate into the SM fermion pair.
For up(down)-type quarks, we obtain
\begin{equation}
	\Gamma(\sigma \to f_{u(d)}\bar f_{u(d)})=\frac{3 \ell^2}{16\pi}\frac{m_\sigma m_f^2}{f_a^2}
	\left( \frac{c_{u(d)} \mu^2}{m_h^2-m_\sigma^2} \right)^2 \left(1- \frac{4m_f^2}{m_\sigma^2} \right)^{3/2},
\end{equation}
where $m_f$ denotes the final state fermion mass and
\begin{eqnarray}
	c_u = \frac{\cos\alpha}{\sin\beta}\left[ \frac{B}{\mu} \cos(\alpha+\beta)+2\sin(\alpha-\beta) \right]
	+\frac{\sin\alpha}{\sin\beta}\frac{m_h^2-m_\sigma^2}{m_H^2-m_\sigma^2}
	\left[ \frac{B}{\mu} \sin(\alpha+\beta)-2\cos(\alpha-\beta) \right],\\
	c_d = -\frac{\sin\alpha}{\cos\beta}\left[ \frac{B}{\mu} \cos(\alpha+\beta)+2\sin(\alpha-\beta) \right]
	+\frac{\cos\alpha}{\cos\beta}\frac{m_h^2-m_\sigma^2}{m_H^2-m_\sigma^2}
	\left[ \frac{B}{\mu} \sin(\alpha+\beta)-2\cos(\alpha-\beta) \right].
\end{eqnarray}
For the decay into charged leptons, we obtain
\begin{equation}
	\Gamma(\sigma \to l\bar l)=\frac{\ell^2}{16\pi}\frac{m_\sigma m_l^2}{f_a^2}
	\left( \frac{c_{d} \mu^2}{m_h^2-m_\sigma^2} \right)^2 \left(1- \frac{4m_l^2}{m_\sigma^2} \right)^{3/2},
\end{equation}
where $m_l$ is the final state lepton mass.

For later convenience, we describe the mixing of CP-odd parts.
They form mass eigenstates as 
\begin{eqnarray}
	G      &=&a_d\cos\beta -a_u \sin\beta,\\
	a_H &=&a_d\sin\beta +a_u \cos\beta +\frac{\ell v\sin(2\beta) }{2f_a} a,\\
	\tilde a   &=&a - \frac{\ell v\sin(2\beta) }{2f_a}(a_d\sin\beta +a_u \cos\beta),
\end{eqnarray}
where $G$ corresponds to the Goldstone boson eaten by the $Z$-boson,
$a_H$ is identified as the CP-odd Higgs boson, and $\tilde a$ is the massless mode in association with
the spontaneous PQ symmetry breaking.

%%%%%%%%%%%%%%%%%%%%%%%%%%%%%%%%%%%%
\subsubsection{Decay into Higgs bosons}
%%%%%%%%%%%%%%%%%%%%%%%%%%%%%%%%%%%%

If the saxion is heavier than the twice the (lightest) Higgs boson mass, the saxion can decay into the
a pair of the Higgs boson.
The decay rate is given by
\begin{equation}
	\Gamma(\sigma \to 2h)=\frac{n^2}{16\pi}\frac{m_\sigma^3}{f_a^2}\left( \frac{\mu}{m_\sigma} \right)^4
	\left(1+ \frac{B\sin(2\alpha)}{2\mu} \right)^2
	\left( 1-\frac{4m_h^2}{m_\sigma^2} \right)^{1/2}.
\end{equation}
%%
%and
%%
%\begin{equation}
%	\Gamma(\sigma \to hH)=\frac{n^2}{16\pi}\frac{m_\sigma^3}{f_a^2}\left( \frac{\mu}{m_\sigma} \right)^4
%	\left(1- \frac{B\cos(2\alpha)}{\mu} \right)^2
%	\left( 1-\frac{4m_h^2}{m_\sigma^2} \right)^{1/2}.
%\end{equation}
%%

%%%%%%%%%%%%%%%%%%%%%%%%%%%%%%%%%%%%
\subsection{$\bar\phi$ decay}
%%%%%%%%%%%%%%%%%%%%%%%%%%%%%%%%%%%%

Next let us estimate the decay rate of $\bar\phi$.
It decays into the axion pair, gauge boson pair and the SM fermions through the $\phi$--$\bar\phi$ mixing.
Roughly speaking, the decay rates of $\bar\sigma$ and $\bar a$ are suppressed
by the mixing factor $\sim (n\bar v /f_a)^2$ compared with the saxion decay rate,
since $\bar\phi$ does not have direct couplings to light particles.

%%%%%%%%%%%%%%%%%%%%%%%%%%%%%%%%%%%%
\subsubsection{Decay into axions}
%%%%%%%%%%%%%%%%%%%%%%%%%%%%%%%%%%%%

The $\bar\sigma$ decay rate into two axions can be read from (\ref{lag}), taking into account the mixing between $a$ and $\bar a$,
as
\begin{equation}
	\Gamma(\bar\sigma\to 2\tilde a) = \frac{(n^2+2n-2)^2}{64\pi}\frac{m_{\bar\sigma}^3}{f_a^2}\left( \frac{\bar v}{f_a} \right)^2.
\end{equation}
%%

%%%%%%%%%%%%%%%%%%%%%%%%%%%%%%%%%%%%
\subsubsection{Decay into gluons}
%%%%%%%%%%%%%%%%%%%%%%%%%%%%%%%%%%%%

From Eq.~(\ref{aGG}), the $\bar\sigma (\bar a)$ decay rate into the gluon pair is obtained as
\begin{equation}
	\Gamma(\bar\sigma\to 2g) \simeq \frac{\alpha_s^2}{32\pi^3}
	\frac{m_{\bar\sigma}^3}{f_a^2} \left(\frac{n^2(n-2)\bar v}{f_a}\right)^2,
\end{equation}
and
\begin{equation}
	\Gamma(\bar a\to 2g) \simeq \frac{\alpha_s^2}{32\pi^3}
	\frac{m_{\bar\sigma}^3}{f_a^2} \left(\frac{n\bar v}{f_a}\right)^2.
\end{equation}
%%

%%%%%%%%%%%%%%%%%%%%%%%%%%%%%%%%%%%%
\subsubsection{Decay into fermions}
%%%%%%%%%%%%%%%%%%%%%%%%%%%%%%%%%%%%

The $\bar\sigma$ decays into fermions through the mixing of $\sigma$--$\bar\sigma$.
From Eq.~(\ref{s-sbarmix}), we find the decay rate into a fermion pair as
\begin{equation}
	\Gamma(\bar\sigma\to f\bar f)\simeq \left(\frac{n^2(n-2)\bar v}{f_a}\right)^2 \Gamma(\sigma\to f\bar f),
\end{equation}
after the $m_\sigma$ in the formula is replaced with $m_{\bar\sigma}$.
Similarly, we can find the decay rate of $\bar a$ into an up(down)-type quark pair from the mixing given in (\ref{a-abarmix}), as
\begin{equation}
	\Gamma(\bar a \to f_{u(d)}\bar f_{u(d)})= \frac{3\ell^2}{16\pi}\frac{m_{\bar a}m_f^2}{f_a^2}
	\left(\frac{c'_{u(d)}n\bar v}{f_a}\right)^2\left(1- \frac{4m_f^2}{m_{\bar a}^2} \right)^{3/2},
\end{equation}
where
\begin{equation}
	c'_u = \frac{(\tan\beta)^{-1}}{\tan\beta+(\tan\beta)^{-1}},~~~c'_d = \frac{\tan\beta}{\tan\beta+(\tan\beta)^{-1}}.
\end{equation}
The decay rate into a charged lepton pair is given by
\begin{equation}
	\Gamma(\bar a \to l\bar l)= \frac{\ell^2}{16\pi}\frac{m_{\bar a}m_l^2}{f_a^2}
	\left(\frac{c'_{d}n\bar v}{f_a}\right)^2\left(1- \frac{4m_l^2}{m_{\bar a}^2} \right)^{3/2}.
\end{equation}
%%

%%%%%%%%%%%%%%%%%%%%%%%%%
\section{Neutrino mass}
%%%%%%%%%%%%%%%%%%%%%%%%%

All gauge-invariant and R-parity conserving operators up to dimension four in the superpotential are listed
in Table.~\ref{table:dim2}-\ref{table:dim4}.
It is seen that the neutrino mass operator $LH_u LH_u$ is allowed for $q_H=-4/5$.
Actually, the operator
\begin{equation}
	W = \frac{1}{2} M_N NN + y_N NLH_u,
\end{equation}
is allowed for $q_H=-4/5$ if the right-handed neutrino has a zero PQ charge.
In this case, the higher dimensional operators 
\begin{equation}
	W = \frac{1}{M_*}\left( QQQ'H_d + Q\bar U'\bar E H_d + Q\bar U\bar E' H_d  \right)
\end{equation}
are also allowed and the vector-like particles can decay ($M_*$ is the cut off scale).
The K\"ahler potential like 
\begin{equation}
	K = \frac{1}{M_*}\left(Q\bar U' L^\dagger+ \bar U\bar E' \bar D^\dagger
	+E'^\dagger H_d H_d + \bar E' H_d H_u^\dagger \right) + {\rm h.c.}
\end{equation}
are allowed, which also induce the decay of vector-like matter. Note that the life time of the proton is  sufficiently long, since its decay width is suppressed by ${M_*}^{-4}$.

%%%%%%%%%%%%%%%% table %%%%%%%%%%%%%%%%%%%%%%
\begin{table}
  \begin{center}
    \begin{tabular}{  | c | c |  c | }
      \hline 
         Operator & Component  & PQ charge \\
       \hline 
        ${\bf 5_H}{\bf \overline{5}_H}$             &  $H_u H_d$ & $-2$ \\
        ${\bf 10'}{\bf \overline{10}'}$                  &  $Q'\bar Q', U'\bar U', E'\bar E'$ & $-2$ \\
      \hline
    \end{tabular}
    \caption{ 
    		R-parity conserving dimension-2 superpotentials.
           }
    \label{table:dim2}
  \end{center}
\end{table}
%%%%%%%%%%%%%%%%%%%%%%%%%%%%%%%%%%%%%%%%%%%%%% 

%%%%%%%%%%%%%%%% table %%%%%%%%%%%%%%%%%%%%%%
\begin{table}
  \begin{center}
    \begin{tabular}{  | c | c |  c | }
      \hline 
         Operator & Component  & PQ charge \\
       \hline 
        ${\bf 10_M}{\bf 10_M}{\bf 5_H}$                  &  $Q\bar U H_u$ & $0$ \\
        ${\bf 10_M}{\bf \bar 5_M}{\bf \bar 5_H}$    &  $Q\bar D H_d, L\bar E H_d$ & $0$ \\
         ${\bf 10'}{\bf 10'}{\bf 5_H}$    &  $Q'\bar U' H_u$ & $0$ \\
         ${\bf \overline{10}'}{\bf \overline{10}'}{\bf \bar 5_H}$    &  $\bar Q' U' H_d$ & $-6$ \\
         ${\bf 10'}{\bf \bar 5_M}{\bf \bar 5_M}$    &  $Q' L\bar D, \bar U'\bar D\bar D, \bar E' LL$ & $4+5q_H/2$ \\
         ${\bf 10'}{\bf \bar 5_H}{\bf \bar 5_H}$    &  $\bar E' H_d H_d$ & $-4-5q_H/2$ \\
         ${\bf \overline{10}'}{\bf 5_H}{\bf 5_H}$    &  $E' H_u H_u$ & $-2+5q_H/2$ \\
      \hline
    \end{tabular}
    \caption{ 
    		R-parity conserving dimension-3 superpotentials.
           }
    \label{table:dim3}
  \end{center}
\end{table}
%%%%%%%%%%%%%%%%%%%%%%%%%%%%%%%%%%%%%%%%%%%%%% 

%%%%%%%%%%%%%%%% table %%%%%%%%%%%%%%%%%%%%%%
\begin{table}
  \begin{center}
    \begin{tabular}{  | c | c |  c | }
      \hline 
         Operator & Component  & PQ charge \\
       \hline 
        ${\bf 10_M}{\bf 10_M}{\bf 10_M}{\bf \bar 5_M}$   &  $QQQL, \bar U\bar U\bar E\bar D, Q\bar U\bar E L, QQ\bar U\bar D$ & $2$ \\
        ${\bf 10'}{\bf 10'}{\bf 10_M}{\bf \bar 5_M}$   &  $Q'Q'QL, \bar U'\bar U'\bar E\bar D, \dots$ & $2$ \\
        ${\bf 10_M}{\bf 10_M}{\bf 10}'{\bf \bar 5_H}$   &  $QQQ'H_d, Q'\bar U\bar E H_d, Q\bar U'\bar E H_d, Q\bar U\bar E' H_d$ & $-2-5q_H/2$ \\
        ${\bf 10'}{\bf 10'}{\bf 10}'{\bf \bar 5_H}$   &  $Q'Q'Q'H_d, Q'\bar U'\bar E' H_d$ & $-2-5q_H/2$ \\
        ${\bf \overline{10}'}{\bf \overline{10}'}{\bf \overline{10}}'{\bf 5_H}$   &  $\bar Q'\bar Q'\bar Q'H_u, \bar Q' U' E' H_u$ & $-6+5q_H/2$ \\
        ${\bf 5_H}{\bf \bar 5_M}{\bf 5_H}{\bf \bar 5_M}$   &  $LH_u LH_u$ & $4+5q_H$ \\
        ${\bf 10_M}{\bf \overline{10}'}{\bf 10_M}{\bf \overline{10}'}$   &  $QQ\bar Q'\bar Q', \bar U\bar UU'U', \bar E\bar E E'E' $ & $-4$ \\
        ${\bf 10'}{\bf \overline{10}'}{\bf 5_H}{\bf \overline{5}_H}$   &  $Q'\bar Q' H_u H_d, U'\bar U' H_u H_d, E'\bar E' H_u H_d$ & $-4$ \\
        ${\bf 10_M}{\bf \overline{10}'}{\bf 5_H}{\bf \overline{5}_M}$   &  $Q\bar Q' LH_u, U'\bar U LH_u, E'\bar E LH_u$ & $5q_H/2$ \\
      \hline
    \end{tabular}
    \caption{ 
    		R-parity conserving dimension-4 superpotentials.
           }
    \label{table:dim4}
  \end{center}
\end{table}
%%%%%%%%%%%%%%%%%%%%%%%%%%%%%%%%%%%%%%%%%%%%%% 

%%%%%%%%%%%%%%%% table %%%%%%%%%%%%%%%%%%%%%%
\begin{table}
  \begin{center}
    \begin{tabular}{  | c | c |  c | }
      \hline 
         Operator & Component  & PQ charge \\
       \hline 
        ${\bf 10_M}{\bf 10_M}{\bf \bar 5_H}^\dagger$                  &  $Q\bar U H_d^\dagger$ & $2$ \\
        ${\bf 10'}{\bf 10'}{\bf \bar 5_H}^\dagger$                  &  $Q'\bar U' H_d^\dagger$ & $2$ \\
        ${\bf 10_M}{\bf 10'}{\bf \bar 5_M}^\dagger$                  &  $Q\bar U' L^\dagger, \bar U\bar E' \bar D^\dagger$ & $-2-5q_H/2$ \\
        ${\bf \overline{10}'}{\bf \overline{10}'}{\bf 5^\dagger_H}$                  &  $\bar Q' U' H_u^\dagger$ & $-4$ \\
        ${\bf {10}'}{\bf \overline{10}'^\dagger}{\bf 5_H}$                  &  $Q' U'^\dagger H_u$ & $2$ \\
        ${\bf {10}'^\dagger}{\bf \overline{10}'}{\bf \bar 5_H}$                  &  $Q'^\dagger U' H_d$ & $-4$ \\
        ${\bf {10}^\dagger_M}{\bf \overline{10}'}{\bf \bar 5_M}$             &  $Q^\dagger U' L, \bar U^\dagger E' \bar D$ & $-5q_H/2$ \\
      \hline
        ${\bf \overline{10}'^\dagger}{\bf \bar 5_M}{\bf \bar 5_M}$   &  $\bar Q'^\dagger L\bar D, U'^\dagger\bar D\bar D, E'^\dagger LL$ & $6+5q_H/2$ \\
        ${\bf \overline{10}'^\dagger}{\bf \bar 5_H}{\bf \bar 5_H}$   &  $E'^\dagger H_d H_d$ & $-2-5q_H/2$ \\
        ${\bf {10}'^\dagger}{\bf 5_H}{\bf 5_H}$                  &  $\bar E'^\dagger H_u H_u$ & $5q_H/2$ \\
        ${\bf \overline{10}'}{\bf 5_H}{\bf \bar 5_H^\dagger}$        &  $E' H_u H_d^\dagger$ & $5q_H/2$ \\
        ${\bf {10}_M}{\bf \bar 5_M}{\bf 5^\dagger_H}$                  &  $Q\bar D H_u^\dagger, \bar E L H_u^\dagger$ & $2$ \\
        ${\bf {10}'}{\bf \bar 5_H}{\bf 5^\dagger_H}$                  &  $\bar E' H_d H_u^\dagger$ & $-2-5q_H/2$ \\
      \hline
    \end{tabular}
    \caption{ 
    		R-parity conserving dimension-3 K\"ahler potentials.
           }
    \label{table:dim3}
  \end{center}
\end{table}
%%%%%%%%%%%%%%%%%%%%%%%%%%%%%%%%%%%%%%%%%%%%%% 

%%%%%%%%%%%%%%%%%%%%%%%%%%%%%
%\bibliographystyle{JHEP}
%\bibliography{GMSB_PQ}

%%%%%%%%%%%%%%%%%%%%%%%%%%%%%%%%%%%%%

%%%%%%%%%%%%%%%%%%%%%%%%%%%%%%%%%%%%%

\end{document}